# Presence of atomic disorder and its effect on magnetic and electronic properties of NiCrGa half Heusler alloy

Madhusmita Baral[1,2*], M. K. Chattopadhyay[2,3], Aparna Chakrabarti[2,4] and Tapas Ganguli[1,2]

[1]*Electron Spectroscopy and Materials Laboratory, Raja Ramanna Centre for Advanced Technology, Indore 452 013, India*

[2]*Homi Bhabha National Institute, Training School Complex, Anushaktinagar, Mumbai 400 094, India*

[3]*Free Electron Laser Utilization Laboratory, Raja Ramanna Centre for Advanced Technology, Indore 452 013, India*

[4]*Theory and Simulations Laboratory, Raja Ramanna Centre for Advanced Technology, Indore 452 013, India*

*madhusmita@rrcat.gov.in

**Abstract:** In this work, polycrystalline NiCrGa half Heusler alloy, which is predicted to be half-metallic ferromagnet from first principles calculations, has been synthesized by arc meting technique and its structural, magnetic as well as the electronic properties have been studied. The measured x-ray diffraction (XRD) pattern shows the signature of a disordered structure. From the magnetization measurements, there is no evidence of ferromagnetic ordering observed in this system down to the lowest temperature studied. Instead, the system shows the signature of an antiferromagnetic ordering at very low temperature. The experimentally observed structural and magnetic properties are found to be significantly different from the theoretically predicted properties of the ordered cubic $C1_b$ structure. To probe the possible disorder present in the system and its effect on the magnetic properties, we have carried out first principles calculations using the spin-polarized-relativistic Korringa-Kohn-Rostoker method (SPR-KKR). Using a combination of XRD, photoelectron spectroscopy, magnetization measurements and first principles calculations, we conclude that NiCrGa has significant amount of atomic disorder. Although, the ordered structure is energetically more stable than the disordered structures, we find that after synthesis, the system tends to stabilize in a disordered structure. With this atomic disorder present in the sample, the ferromagnetic ordering is disturbed and the calculated spin polarization is consequently reduced.

**INTRODUCTION**

Half-metallic ferromagnets have a metallic band for electrons in one spin channel and a bandgap in the other spin channel, thereby leading to 100% spin polarization at the Fermi level ($E_F$). Because of this property, these materials can be efficient spin injectors for applications in spin-polarized electronics [1]. The first materials predicted to be half metallic ferromagnets were the half-Heusler alloys NiMnSb and PtMnSb, which crystallize in cubic $C1_b$ structure [2-6]. In the ordered half Heusler alloy (XYZ), with $C1_b$ structure (space-group F-43m), the Z, Y and X atoms occupy the A, C and B sites with Wyckoff positions *4a, 4b and 4c* respectively. The site D with Wyckoff position *4d* remains unoccupied. In literature, NiMnSb has been reported to exhibit high Curie temperature [7] and possess an integral magnetic moment of 4.0 $\mu_B$ per unit cell [8]. Due to these properties, there has been interest in the fabrication of tunnel junctions [9, 10] and spin valves [11, 12] using thin films of NiMnSb. However, the measured spin polarization was found to be lower than the predicted value of 100%. Soulen *et al* have measured the spin polarization of NiMnSb thin film, deposited by co-evaporation, by point contact Andreev reflection (PCAR) measurement and found a value only up to 58% [13]. The neutron diffraction measurements on bulk NiMnSb (powder) showed an atomic disorder of ~10% [8]. From x-ray diffraction (XRD) measurement, Kautzky *et al* showed that atomic disorder is significant in the thin films of PtMnSb which causes the reduction of spin polarization [14]. Using first-principles electronic structure calculations, the disorder in NiMnSb has been studied by Orgassa *et al* [15, 16]. They have taken into account three different types of disorders: (a) intermixing of Ni and Mn atoms, (b) migration of Ni and Mn atoms to the unoccupied site, and (c) migration of Mn and Sb atoms to the unoccupied site. They found that 5% of disorder is enough to create impurity states in the minority-spin gap which significantly reduces the spin polarization and the magnetic moment. Galanakis *et al* have studied the spin-polarization and electronic properties of various half-metallic Heusler alloys using first principles calculations [17, 18]. They showed that the factors such as changes in the lattice parameter, spin–orbit coupling, doping, disorder and defects affect the half-metallicity in the Heusler alloys. Even in the case of the half-metallic ferromagnet PtMnSb, it has been reported that atomic disorder can significantly influence the half-metallicity of the material [4]. Similarly, another alloy IrMnGa, which is antiferromagnetic at low temperatures ($T_N$ = 60 K), has been shown to exhibit high atomic disorder from the neutron diffraction measurement [8]. In this compound, there is a mixing of Ir at (000) and Mn at (1/4 1/4 1/4) locations. While 62.6% of Ir occupies its original A site (0 0 0), the remaining 37.4% Ir is present at the site B (1/4 1/4 1/4). Thus atomic disorder plays an important role in reducing the spin polarization and magnetic moment in these materials.

Recently, Luo *et al* have studied the electronic structure and magnetic properties of the half-Heusler alloys NiCrAl, NiCrGa and NiCrIn using the self-consistent full-potential linearized-augmented plane wave (FLAPW) method based on the local spin density approximation (LSDA) within the density functional theory (DFT) [19]. These alloys are predicted to be half metals (100% spin polarization) with a total magnetic moment of 1 μB/f.u. (which agree with the Slater–Pauling rule) and the estimated half metallic gap of NiCrAl, NiCrGa and NiCrIn are found to be 0.02 eV, 0.17 eV and 0.07 eV respectively. In these alloys the Cr atom carries a large magnetic moment and is antiferromagnetically aligned to Ni and Al (Ga and In). Although, there is an extensive amount of work on the Mn based half-Heusler half-metallic alloys, experimental studies on the NiCrZ half Heusler alloys (where Z = Al, Ga, In) are not found in the literature.

As the reported half metallic gap in NiCrGa is the highest among the above three [19], the aim of the present work is to synthesize the alloy and experimentally probe its half metallic ferromagnetic character through the study of structural, magnetic and electronic properties. The synthesized NiCrGa half Heusler alloy has been found to possess a significant amount of atomic disorder, and the experimentally observed magnetic property is found to be significantly different from the reported calculated property. To understand our experimental results, we have carried out theoretical calculations by taking into account different types of disorders in the structure. From the combination of experimental and theoretical studies, we have identified the nature of atomic disorder present in the NiCrGa sample. For all the possible disorders present in the sample, the calculated spin polarization in each case is found to be rather small.

**EXPERIMENTAL AND COMPUTATIONAL DETAILS**

Polycrystalline ingot of NiCrGa half Heulser alloy with nominal composition $Ni_{33.3}Cr_{33.3}Ga_{33.3}$ has been synthesized in an arc-melting furnace under argon atmosphere. After synthesis, the sample was annealed in vacuum ($8 \times 10^{-8}$ mbar) at 800℃ for 3 days and then cooled down to room temperature by switching off the power to the furnace. The bulk composition of the sample was found to be $Ni_{33.2}Cr_{35.3}Ga_{31.5}$ by x-ray fluorescence (XRF) measurements performed using the X-ray fluorescence microprobe beamline (BL-16) of Indus-2. Hereafter, the sample shall be referred to as NCG. The structural characterization of the sample was performed with the help of room temperature x-ray diffraction (XRD) measurements, using a Bruker Discover system with a Cu Kα x-ray source (λ=1.5405 Å). In the ordered $C1_b$ structure, the XRD pattern contains three types of peaks corresponding to: (a) fundamental reflections where *h, k, l* are all even and $h + k + l = 4n$ (n=1,2,3,4…), (b) type-I superlattice reflections where $h, k, l$ are all odd and (c) type-II superlattice reflections where $h, k, l$ are all even and $h + k + l = 2n$ (n=1,3,5,7…). The temperature ($T$) and field ($H$) dependence of magnetization ($M$) were measured using a superconducting quantum interference devise based vibrating sample magnetometer (MPMS-3 SQUID VSM, Quantum Design). The photoelectron spectroscopy (PES) measurements were carried out using a PHOIBOS 150 hemispherical analyzer with a twin anode Al Kα (1486.6 eV)/MgKα (1253.6 eV) X-ray source and He-I UV source (21.218 eV). The vacuum inside the analysis chamber during the experiment was $8\times10^{-11}$ mbar. Before the PES measurements, the sample was cleaned in-situ by sputtering with 2 keV Argon ions at an ion current of 10 mA. This was followed by annealing at 500 K. The surface composition was confirmed to be close to the bulk value.

To understand the experimental results, we have calculated the magnetic moment and density of states (DOS) for the ordered and different types of disordered structures of NCG using the spin-polarized-relativistic Korringa-Kohn-Rostoker method (SPR-KKR) in combination with the full potential spin-polarized scalar-relativistic (SP-SREL) mode [20]. For the exchange-correlation potential, the Generalized Gradient Approximation (GGA (PBE)) has been used for all the calculations. The self-consistent field (SCF) calculations have been performed with the tolerance for energy convergence as $10^{-5}$ $Ry$. The mesh of k points for the SCF cycles has been taken as 21×21×21 in the Brilloiun Zone (BZ) and the number of E-mesh points was taken as 100. The angular momentum expansion up to $l_{max} = 3$ has been taken for each atom.

**RESULTS AND DISCUSSION**

**Structural properties:**

The measured room temperature XRD pattern for the present sample is shown in figure 1(a) where the peaks are indexed for the cubic $C1_b$ structure. The lattice parameter calculated from these peaks is *a = b = c = 5.80* Å, which is approximately 5% higher than the equilibrium lattice parameter of the ordered $C1_b$ structure calculated by us using the Vienna Ab initio Simulation Package (VASP) and by H. Luo *et al* [19]. In the XRD pattern, there is no signature of the type-I superlattice reflection peaks and the intensity of type-II superlattice peaks is negligibly small (almost merged in the background). This clearly indicates the presence of atomic disorder in the sample. To identify the type of disorder present in the sample, we have simulated the XRD patterns of many possible atomic disordered configurations (by considering various types of intermixing of the atoms and site occupancy of the atoms at different sites), using "Powder cell" software [21]. The types of disorders which are more relevant are discussed here. The types of disorder and the site occupancies corresponding to these disorders are summarized in table 1 and the simulated XRD patterns are shown in figure 1(b). Type-0 is ~~for~~ the ordered case with the ideal $C1_b$ structure. In the type-1 disordered case, there is an intermixing of Cr and Ga atoms and the vacant D site is partially occupied by the Ni atom. The type-2 has only an intermixing of Cr and Ga atoms. In type-3, the vacant D site is occupied by Cr and Ga. In type-4, the unoccupied D-site is occupied by Ni, Cr and Ga. Type-5 is the case where the site D is partially occupied by Ni. In type-6, the vacant site is occupied by Ni and Cr. In all the cases the stoichiometry of Ni:Cr:Ga was kept constant at 1:1:1. Some of the disorders considered in this work have also been previously reported by others for other materials [15, 16].

***Table-1:*** *The first column shows the types of disorder considered. The second column shows the sites with interchange of atoms (or vacancies). Occupancies of the four lattice sites are shown for the different types of disorder. Simulation has been carried out with experimental lattice parameter a=b=c=5.80 Å and space-group F-43m (216). In the half Heusler alloy XYZ with $C1_b$ structure, the Z, Y and X atoms occupy the A, C and B sites with Wyckoff positions 4a (0 0 0), 4b (1/2 1/2 1/2) and 4c (1/4 1/4 1/4) respectively and the site D with Wyckoff position 4d (3/4 3/4 3/4) is unoccupied.*

| Disorder | | Site occupancies | | | |
|---|---|---|---|---|---|
| Type | Scheme | A | C | B | D |
| Type-0 | | Z | Y | X | ---- |
| Type-1 | A↔C  B→D | $Z_{0.5} Y_{0.5}$ | $Y_{0.5} Z_{0.5}$ | $X_{0.5}$ | $X_{0.5}$ |
| Type-2 | A↔C | $Z_{0.5} Y_{0.5}$ | $Y_{0.5} Z_{0.5}$ | X | ---- |
| Type-3 | AC →D | $Z_{0.5}$ | $Y_{0.5}$ | X | $Z_{0.5} Y_{0.5}$ |
| Type-4 | ABC →D | $Z_{0.75}$ | $Y_{0.75}$ | $X_{0.75}$ | $Z_{0.25} Y_{0.25} X_{0.25}$ |
| Type-5 | B→ D | Z | Y | $X_{0.5}$ | $X_{0.5}$ |
| Type-6 | BC →D | Z | $Y_{0.6}$ | $X_{0.6}$ | $X_{0.4} Y_{0.4}$ |

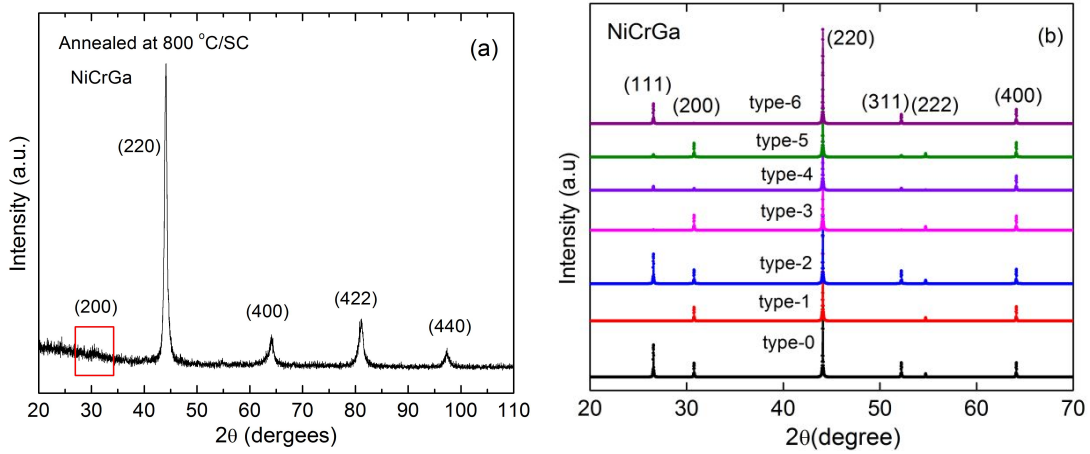

*Figure 1: (a) Room temperature XRD patterns of NiCrGa annealed at 800°C followed by slow cooling (SC). (b) Simulated XRD patterns of NCG using Cu kα x-ray energy and the experimental lattice parameters (5.8 Å) for different types of disorders.*

The intensities of the lattice reflections are proportional to the square of the structure factor, which is of the form: $F_{hkl} = \sum_{j=1}^{n} f_j Exp\{2\pi i(hx_j + ky_j + lz_j)\}$, where $f_j$ is the *form factor* of a particular atom. The $|F_{hkl}|^2$ of the fundamental peaks is $16(f_{Ga} + f_{Cr} + f_{Ni})^2$ for both the ordered and all the disordered systems, whereas the $|F_{hkl}|^2$ (intensity) of the superlattice peaks depends on the type of disorder. Therefore, different types of atomic disorder can be distinguished by comparing the intensity ratios of the superlattice and fundamental reflections of one or more of the following: I(111)/I(220), I(200)/I(220) and I(200)/I(400). The calculated ratio of intensities I(111)/I(200), I(200)/I(220) and I(200)/I(400), for the different types of disorder, are shown in table 2. In the experimentally observed XRD pattern, the (111) reflection is absent and the intensity of the (200) reflection is significantly smaller than the fundamental (220) and (400) reflections. Analyzing the values of the various intensity ratios mentioned in table-2, it is found that the possible atomic disorder can be of types 1, 3 and 5, where the intensity of (111) is zero or negligible with respect to that of (200). The disorder of the types 2 and 6 are clearly ruled out, as in these cases the intensity of the (111) reflection is significantly higher than the (200) reflection. Regarding the disorder of type 4, both the (111) and (200) reflections are of very low intensity, with the (111) reflection having slightly higher intensity with respect to the (200) reflection. Hence this type of disorder may also be considered to be possible here. Although we have rotated the sample during the measurement, the presence of some orientational effects cannot really be ruled out, and hence the intensity ratio alone cannot be used for the determination of the actual disorder type and absolute disorder fraction, present in the sample.

*Table-2:* *Calculated and experimental intensity ratios of super-lattice and fundamental peaks.*

| Disorder | $\dfrac{I(111)}{I(200)}$ | $\dfrac{I(111)}{I(220)}$ | $\dfrac{I(200)}{I(220)}$ | $\dfrac{I(200)}{I(400)}$ | $\dfrac{I(400)}{I(220)}$ |
|---|---|---|---|---|---|
| Type-0 | 2.34 | 0.34 | 0.14 | 0.96 | 0.15 |
| Type-1 | 0 | 0 | 0.14 | 0.96 | 0.15 |
| Type-2 | 2.17 | 0.31 | 0.14 | 0.96 | 0.15 |
| Type-3 | 0.04 | 0.01 | 0.16 | 1.06 | 0.15 |
| Type-4 | 6.30 | 0.03 | 0.00 | 0.00 | 0.15 |
| Type-5 | 0.17 | 0.02 | 0.14 | 0.96 | 0.15 |
| Type-6 | 90.30 | 0.21 | 0.00 | 0.02 | 0.15 |
| Experiment | 0 | 0 | 0.01-0.02 | 0.20±0.05 | 0.09±0.01 |

From the XRD measurement, the experimental lattice parameter is found to be approximately 5% higher than the calculated equilibrium lattice parameter of the ordered structure of NCG. To probe this difference, we have determined the equilibrium lattice parameter for the ordered as well as for all the possible types of atomic disorders identified in the XRD measurements (type 1, 3, 4 and 5) by calculating the total energy of the system as a function of lattice parameter, using the SPR-KKR method. In figure 2, we have plotted the difference of total energy with respect to the minimum energy ($\Delta E$) as a function of the lattice parameter for the ordered as well as the type-1 and type-3 disordered cases (the change of $\Delta E$ as a function lattice parameter for type-4 and type-5 disorders is similar to that observed in types-1 and 3 disorders, hence not shown here). The $\Delta E$ shows a minimum at 5.5 Å for the ordered structure (type-0) and at about 5.9 ± 0.05 Å for all the possible types of disorders studied in the present work. The equilibrium lattice parameter of the disordered structures is quite close to the experimentally observed value of 5.8 Å. Although energetically the ordered structure is more stable than the disordered structures, we find that after synthesis, the system tends to stabilize in a disordered structure with an increased lattice parameter.

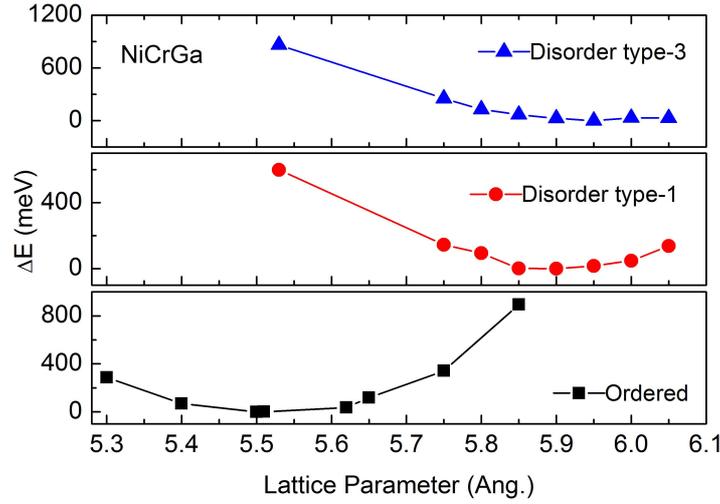

*Figure 2*: *Difference of total energy with respect to the minimum energy (ΔE) plotted as a function of lattice parameter for the ordered as well as the type-1 and type-3 disordered cases.*

**Magnetic and Electronic properties:**

The $M(T)$ curve for the present sample between 2 K to 250 K, for $H$ = 100 Oe, is shown in figure 3 (a). From the figure, it is clear that the sample is paramagnetic (PM) at high temperature. The field cooled cooling (FCC) $M(T)$ curve for $H$ = 100 Oe has an inflection point at around 5.4 K (see inset (i) of figure 3(a)), which is located with the help of the derivative curve shown in the inset (ii) of figure 3(a). The presence of such an inflection point generally indicates a PM to ferromagnetic (FM) phase transition in the sample. In higher fields (50 kOe), the inflection point shifts to higher $T$ (14 K, see the inset (iii) of figure 3(a)). This indicates that there are short range magnetic correlations in the sample in this $T$-regime, and the application of magnetic field further strengthens these correlations. However, below 4.5 K, the FCC $M(T)$ curve for $H$ = 100 Oe shows a decrease of $M$ with decreasing $T$, before a second increase of $M$ below 3.2 K. This behavior is unlike that of a FM system, and this is consistent with the isothermal $M(H)$ curves shown in figure 3(b). While there is strong non-linearity in the low $H$ region of the $M(H)$ curves at low *T*, the Arrott plots [22] (inset to figure 3 (b): M² versus H/M), confirm the absence of spontaneous magnetization. Hence, the sample is not FM at any *T*. The non-linearity of the *M(H)* curves at low *T* and at low H regime, however, indicates the presence of short range magnetic correlations in the sample. It may be noted that though the non-linearity in the *M(H)* curves becomes less significant with increasing *T*, even at 15 K, which is close to the inflection point in the *M(T)* curve for *H* = 50 kOe, the *M(H)* curves are appreciably nonlinear in low *H* (below 10 kOe).

From the *M(T)* curve for *H* = 100 Oe, we have obtained the $\chi$ versus *T* curve for the present sample, which is shown figure 3(c). We have found that this curve can be best fitted with the following expression:

$$\chi = \chi_0 + \frac{C_{Curie}}{(T + T_N)} \quad (1)$$

Here, $\chi$ is the susceptibility, $\chi_0$ is the temperature independent term, $C_{Curie}$ is the Curie constant and $T_N$ is the *Neel* temperature. The temperature independent term $\chi_0$ in the above equation is probably contributed by the conduction electrons [23]. From the Curie-Weiss term (the 2nd term on the right hand side of equation (1) in the above equation, we find the value of $T_N$ to be 7.1 K. From the above fitting, the effective moment comes out to be 0.04 $\mu_B$ which is much smaller than the theoretical value of about 1 $\mu_B$/f.u. reported in the literature [19]. Thus, the reduction of *M* with decreasing *T* below 4.5 K on the FCC *M(T)* curve, which gives rise to a peak at this *T*, may be because of an antiferomagnetic (AFM) ordering. Evidently, the magnetic behavior of the sample is unlike the predicted FM ordering of the ordered NCG [19], and this may be due to the atomic disorder present in the material. It is interesting to note here that, from an ab-initio calculations on Ni$_2$CrGa, it has been observed that the alloy is likely to possess an AFM configuration [24].

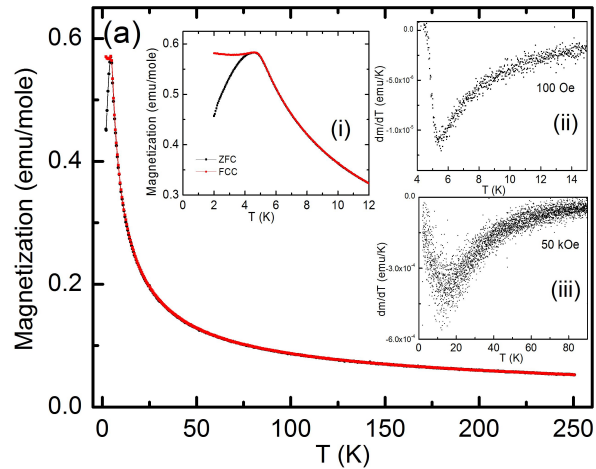

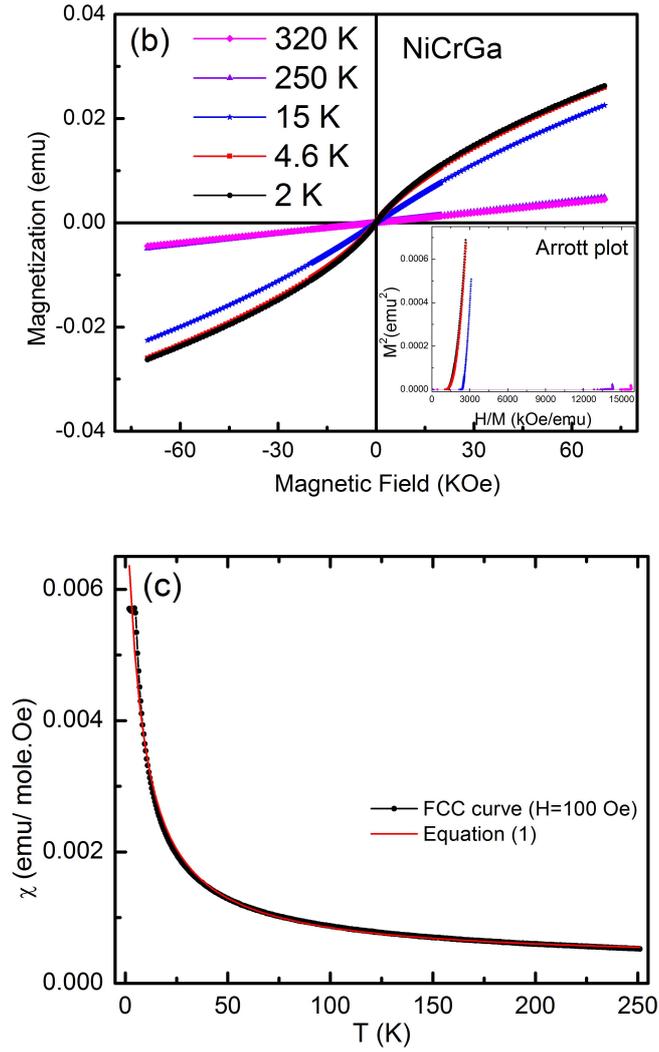

***Figure 3:*** *(a) The M versus T curve in 100 Oe magnetic field, from 0 to 250 K. Inset (i): The M versus T curve in 100 Oe magnetic field at low temperatures. Insets (ii) and (iii): The dM/dT versus T curves in 100 Oe and 50 kOe magnetic fields respectively. (b) The isothermal M versus H curves at different temperatures. The inset shows the Arrott plots. (c) The temperature dependence of DC susceptibility. The dots represent the experimental data, and the fitted line represents equation (1).*

In order to understand the above results, we now look into the calculated (using the experimental lattice parameter of 5.8 Å) total and partial spin magnetic moments for all the possible types of atomic disorders identified from the XRD measurements (type 1, 3, 4 and 5), along with the ordered case. These results are summarized in table 3. We observe that,

(a) for the ordered structure, Cr atom carries a large spin moment of 3.32 $\mu_B$ and the spin

moments of Ga, Cr and Ni are parallel to each other. This leads to a large total moment of 3.42 $\mu_B$ for the experimental lattice parameter. The calculated total magnetic moment is much larger than what is observed experimentally.

(b) In the case of type-1 disorder, the Ni atoms at two different sites (B and D) carry small spin moments of equal magnitude. The Ga atoms at the sites A and C carry small spin moments, but they are anti-parallel to each other. Similarly the Cr atoms at the sites A and C are anti-parallel to each other, and carry a large moment of almost equal magnitude (~3.50 $\mu_B$). This gives a total magnetic moment of 0.02 $\mu_B$/ f.u.

(c) In type-3 disorder, the moments on the Cr atoms at two different sites (C and D) are anti-parallel to each other. The Cr atom at the C site is antiparallel to the Ga atom sitting at the two sites and parallel to the Ni atom (B site) which carries a small positive spin moment. The total magnetic moment in this type of disorder is very small (0.32 $\mu_B$/ f.u.). The total moment in both the disorder cases (types 1 and 3) are close to the measured experimental value but very small compared to the calculated value of 1 $\mu_B$/ f.u. for the ordered structure [19].

(d) For type-4 disorder, the total magnetic moment is 1.84 $\mu_B$/ f.u., where the Cr at site C is anti-parallel to Ga (at its original site A) and parallel to the Ni atom (at its original site B). Moreover, the moment of the Cr atoms sitting at C and D sites are anti-parallel to each other. This value of total magnetic moment is significantly higher than the experimentally observed value.

(e) For the type-5 disorder case, the total moment is found to be 3.80 $\mu_B$/ f.u. This value is also quite large compared to the experimentally observed value.

Thus, from these calculated moments, it seems possible that the present sample has contributions from either one or a combination of the type-1, type-3 and type-4 disordered structures.

**Table-3:** *The calculated total moment and the partial moments at different sites, for different types of disorder in NCG. During the SCF run of the ordered structure, the starting moments on Ni, Cr and Ga were taken as 0.2 $\mu_B$, -1.5 $\mu_B$ and 0.0 $\mu_B$, respectively. For the disordered configurations, the starting partial moments on Ni and Cr, sitting at two different sites, were taken as equal in magnitude and opposite in direction. After SCF convergence, the final moments are found be different and are tabulated below.*

| Types of disorders | Total moment | Partial moment ($\mu_B$) (A-site) | Partial moment ($\mu_B$) (C-site) | Partial moment ($\mu_B$) (B-site) | Partial moment ($\mu_B$) (D-site) |
|---|---|---|---|---|---|
| **Ordered type-0** | $\mu_T$ | $\mu_{Ga}$ | $\mu_{Cr}$ | $\mu_{Ni}$ | -- |
| | 3.42 | 0.00 | 3.32 | 0.10 | -- |
| **Disorder type-1** | $\mu_T$ | $\mu_{Ga}$, $\mu_{Cr}$ | $\mu_{Cr}$, $\mu_{Ga}$ | $\mu_{Ni}$ | $\mu_{Ni}$ |
| | 0.02 | 0.00, -3.50 | 3.46, -0.01 | -0.01 | -0.01 |
| **Disorder type-3** | $\mu_T$ | $\mu_{Ga}$ | $\mu_{Cr}$ | $\mu_{Ni}$ | $\mu_{Cr}$, $\mu_{Ga}$ |
| | 0.32 | 0.04 | -2.39 | -0.24 | 3.31, 0.02 |
| **Disorder type-4** | $\mu_T$ | $\mu_{Ga}$ | $\mu_{Cr}$ | $\mu_{Ni}$ | $\mu_{Cr}$, $\mu_{Ga}$, $\mu_{Ni}$ |
| | 1.85 | -0.04 | 3.13 | 0.22 | -2.87, -0.01, 0.28 |
| **Disorder type-5** | $\mu_T$ | $\mu_{Ga}$ | $\mu_{Cr}$ | $\mu_{Ni}$ | $\mu_{Ni}$ |
| | 3.80 | -0.03 | 3.49 | -0.22 | 0.22 |

To further confirm the type of disorder, we have calculated the total DOS for the type 1, type 3 and type 4 disordered structures and compared them with the experimentally obtained valence band (VB) spectrum. The comparison is shown in figure 4, where the VB spectrum shows three major features near -0.5, -1.3 and -2.2 eV binding energy (BE) values. In the case of type-1 disorder (shown in figures 4 and 5 (a)), the total DOS exhibits two major features near -0.5 and -2.3 eV BE values. The peak near -0.5 eV has a large contribution from the Cr 3d states and a very small contribution from the Ni 3d states, whereas the peak near -2.3 eV has a large contribution from the Ni 3d states and a small Cr 3d state contribution. In contrast to a dip observed near -1.5 eV in the calculated DOS, there is a strong peak in the experimental VB spectrum at the same BE value. From the overall comparison, we find that the DOS of type-1 disorder does not match with the experimental VB spectrum. In type-3 disorder (shown in figures 4 and 5(b)), the main peak of the total DOS exhibits four features near -0.2, -0.6, -1.6 and -2.0 eV BE values. All the features have contributions from Cr and Ni 3d states which are hybridized with each other. In type-4 disorder (shown in figures 4 and 5(c)), the main peak of the total DOS exhibits three features near the BE values of -0.6, -1.6 and -2.3 eV. These features also correspond to the hybridized Cr and Ni 3d states. The overall shapes of the total DOS of type-3 and type-4 disorders match well with our experimental VB spectrum. Only the peak positions are marginally shifted with respect to each other (by ~0.2 eV). Thus, from the comparison of the electronic structure, we infer that type-3 and type-4 are the two possible disorders present in our sample.

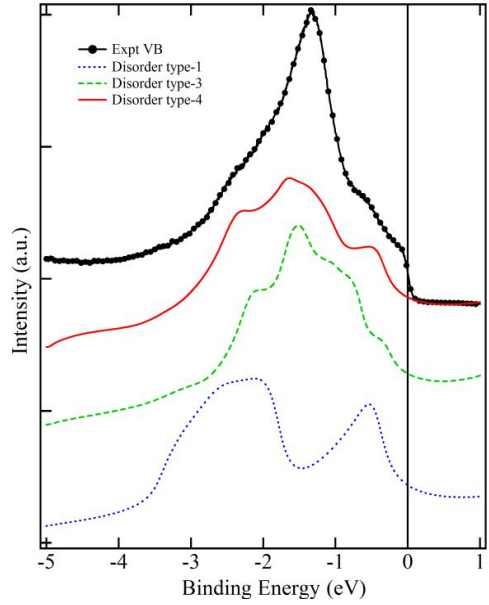

*Figure 4:* *Comparison of experimental VB spectrum with the calculated total DOS for the type-1, type-3 and type-4 disorders of NCG.*

To understand the effect of disorder on the half metallic property of NCG, we have calculated the spin polarization at $E_F$ from the up and down spin total DOS of the ordered structure as well as for the type-3 and the type-4 disordered structures. The spin polarized total DOS for the above mentioned structures are shown in figure 5 (d). For the case of ordered structure, at the equilibrium lattice parameter (5.5 Å), the spin down total DOS exhibits a gap at the $E_F$ and a large spin up DOS corresponding to the Ni and Cr 3d states (at the $E_F$). Thus, at the equilibrium lattice parameter, where the total energy shows the minimum, NCG shows half metallic behavior (100% spin polarization). However, at the experimental lattice parameter (5.8 Å), the spin down gap at the $E_F$ vanishes, and NCG behaves like a metal. In this case the calculated spin polarization is significantly reduced to ~10%. For the type-3 disordered structure, a large spin up and spin down DOS appear at the $E_F$ and the spin polarization has been calculated to be ~1% (at 5.8 Å). Similarly for type-4 disorder, the spin polarization is calculated to be ~13%. Thus, the type-3 and type-4 disorders (which are the possible disorders present in our sample) completely destroy the half metallic character in NCG.

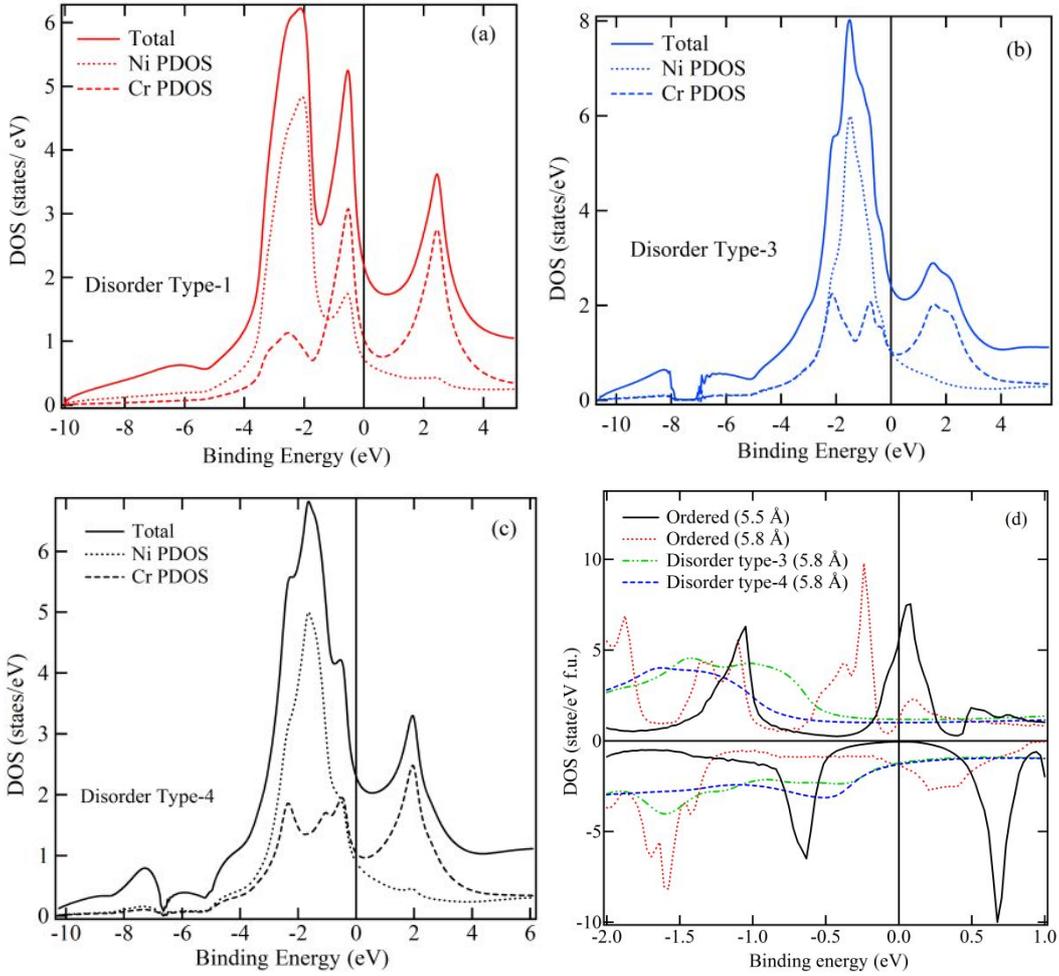

*Figure 5:* *Calculated total DOS and PDOS for the (a) type-1, (b) type-3 and (c) type-4 disorders of NCG. (d) shows the spin polarized total DOS of the ordered (calculated at 5.5 Å and 5.8 Å) and type-3 and type-4 disordered structures (calculated at 5.8 Å).*

Summarizing the results of XRD, $M(T)$, $M(H)$ and PES measurements along with the first principles calculations, we conclude that the NCG sample forms a disordered structure with a possible contribution from type-3 and type-4 disordered configurations described above, and this probably leads to the signature of AFM ordering observed in the *M*(*T*) curves at low *T*. We, however, argue that the short range magnetic correlations (favoured by applied *H*) observed in the sample at *T* above the AFM ordering are of FM nature. This feature, and the inflection point on the *M*(*T*) curves above the AFM ordering temperature, are due to some FM correlations probably contributed by the type-4 disorder, where a significant magnetic moment is expected. Some contribution from the ideal ordered phase with very small concentrations, that go undetected in an XRD measurement, can also not be ruled out as the origin of this FM correlation.

The above results in conjunction with the other reported experimental results, lead us to some important conjectures on the half Heusler based half metallic systems in general. Although predicted to be half metallic from the first principles DFT based calculations, the experiments show that almost all the systems studied so far do not exhibit the predicted 100% spin polarization at $E_F$. Some of these systems include NiMnSb, PtMnSb and IrMnGa [4, 8, 14]. In all these cases, the reduced spin polarization at $E_F$ is attributed to atomic disorder, though the percentage of disorder is found to be different for different compounds. Our work also shows that the presence of a significant amount of atomic disorder in the sample, destroys the predicted FM ordering and half metallicity in NCG. Possibly, the presence of unfilled sites, which are just half a lattice parameter away from each Ni atom, and the similar atomic sizes of Ni, Cr and Ga (Ni:124 pm, Cr:128 pm, Ga:135 pm [25]), makes the system susceptible to disorder, specifically because of the high temperatures, at which it is synthesized. Using a combination of XRD, magnetic and PES measurements as well as first principles calculations, we determine the possible disordered configurations that can be present in NCG system. Our studies highlight the fact that any departure from the crystalline order results in a reduced spin polarization at $E_F$, thereby destroying the half metallic character. Thus, the half Heusler based half metallic systems reported till now do not seem to be promising candidates for applications in spintronics, primarily due to the twin factors of a strong tendency to form in the atomic disordered structure and the strong dependence of the magnetic properties and the half metallic character on the atomic disorder.

**SUMMARY AND CONCLUSION**

NiCrGa half Heusler alloy has been studied experimentally to probe into the theoretically predicted half metallicity and ferromagnetism in this system. Magnetic measurements show no evidence of long range ferromagnetic ordering in this system down to the lowest temperature studied. In fact below 4.5 K, the system shows the signature of an antiferromagnetic ordering. Using a combination of XRD, photoelectron spectroscopy, magnetization measurements and first principles calculations, we conclude that NiCrGa is likely to possess a significant amount of atomic disorder, primarily of the following kinds: (a) type-3 where, the vacant D site is occupied by of the 50% Cr and Ga atoms and (b) type-4, where the vacant D site is occupied by 25% of the Cr, Ga and Ni atoms. Due to these disordered structures, the magnetic order and hence the half metallic property of the system is significantly disturbed, which eventually limits its potential for device applications. Such observations of reduction of spin polarization at $E_F$ from the expected 100% value has been observed in almost all the other half Heusler systems, which have been predicted to be half metals from the theoretical calculations. In almost all the cases studied so far, the presence of atomic disorder has been ascertained experimentally, and this disorder is attributed to the absence of half metallicity. The calculations carried out by us for a large number of disordered configurations clearly show that the half metallic character of NiCrGa system is destroyed for all the atomic disorders studied in this work. Thus from the present study, in conjuction with the available literature, it can be conjectured that the half Heusler alloys may not be the most suitable candidates for half metallicity, due to the possibility of the inherent presence of atomic disorders in the samples.

## ACKNOWLEDGMENTS

The authors wish to thank Dr. P. A. Naik, Director RRCAT and Dr. Arup Banerjee for constant encouragement and support. Dr. Archna Sagdeo is thanked for carrying out XRD measurement, Mr. Ajit Kumar Singh and Mr. Ajay Khooha are thanked for carrying out XRF measurement. Mr. Himanshu Srivastava is thanked for his help during XRF data analysis. Glass blowing facility, RRCAT is thanked for ampoule preparation and vacuum sealing of the sample. MB and AC thank Computer Centre, RRCAT for computer resources.

## REFERENCES

1. G. Prinz, *Phys. Today* **48** (4) (1995) 58

2. R. A. de Groot, F. M. Mueller, P. G. van Engen, and K. H. J.Buschow, *Phys. Rev. Lett.* **50** (1983) 2024

3. E. Kulatov and I. I. Mazin, *J. Phys.: Condens. Matter* **2** (1990) 343

4. H. Ebert and G. Schutz, *J. Appl. Phys.* **69** (1991) 4627

*5*. Xindong Wang, V. P. Antropov, and B. N. Harmon, *IEEE Trans.Magn.* **30** (1994) 4458

*6*. S. J. Youn and B. I. Min, *Phys. Rev. B* **51** (1995) 10436

*7*. M. J. Otto, R. A. M. van Woerden, P. J. van der Valk, J. Wijngaard, C. F. van Bruggen and C. Haas, *J. Phys.: Condens. Matter* **1 (**1989) 2341

*8*. R. B. Helmholdt, R. A. de Groot, F. M. Muller, P. G. van Engen and K. H. J. Buschow, *J. Magn. Magn. Mater.* **43 (**1984) 249

*9*. R. Kabani, M. Terada, A. Roshko, and J. S. Moodera, *J. Appl. Phys*. **67** (1990) 4898

*10*. C. T. Tanaka, J. Nowak, and J. S. Moodera, *J. Appl. Phys.* **81** (1997) 5515

*11*. C. Hordequin, J.P. Nozieres, and J. Pierre, *J. Magn. Magn. Mater.***183** (1998) 225

*12*. J. A. Caballero, Y. D. Park, J. R. Childress, J. Bass, W. -C. Chiang, A. C. Reilly, W. P. Pratt and J. A. Petroff Fero, *J. Vac. Sci. Technol. A* **16 (**1998) 1801

*13*. R. J. Soulen Jr., J. M. Byers, M. S. Osofsky, B. Nadgorny,T. Ambrose, S. F. Cheng, P. R. Broussard, C. T. Tanaka, J. Nowak, J. S. Moodera, A. Barry and J. M. D. Coey, *Science* **282** (1998) 85

*14*. M. C. Kautzky, F. B. Mancoff, J.-F. Bobo, P. R. Johnson, R. L. White, and B. M. Clemens, *J. Appl. Phys*. **81** (1997) 4026

*15*. D. Orgassa, H. Fujiwara, T. C. Schulthess, and W. H. Bulter, *Phys. Rev. B* **60** (1999) 13237

*16*. D. Orgassa, H. Fujiwara, T. C. Schulthess, and W. H. Bulter, *J. Appl. Phys.* **87** (2000) 5870

*17*. I Galanakis and Ph Mavropoulos , *J. Phys.: Condens. Matter* **19** (2007) 315213

*18*. T. Block, M. J. Carey, B. A. Gurney and O. Jepsen, *Phys. Rev. B* **70** (2004) 205114

*19*. H. Luo, Z. Zhu, G. Liu, S. Xu, G. Wu, H. Liu, J. Qu, Y. Li, *Physica B* **403** (2008) 200

*20*. Ebert H et al The Munich SPRKKR package, version 5.4 and 6.3 (http://olymp.cup.unimuenchen.de/ak/ebert/SPRKKR)

*21*.http://www.ccp14.ac.uk/tutorial/powdcell/basic.html

*22*. A. Arrott and J. E. Noakes, *Phys. Rev. Lett.* **19** (1967) 786

*23*. Stephen Blundell, *Magnetism in condensed Matter, Oxford Master series in Condensed Matter Physics.*

24. Tufan Roy and Aparna Chakrabarti, *J. Magn. Magn. Mater.* **401** (2016) 929

25. https://www.ptable.com/

24. Tufan Roy and Aparna Chakrabarti, *J. Magn. Magn. Mater.* **401** (2016) 929

25. https://www.ptable.com/